\documentclass{pasj00}
\draft

\begin{document}
\SetRunningHead{S.Bessho and T.Tsuribe}{Collapse of Filament under UV}

\title{Collapse of Primordial Filamentary Clouds under Far-Ultraviolet Radiation}

\author{Shinji \textsc{Bessho} and Toru \textsc{Tsuribe}}
\affil{Department of Earth and Space Science, Osaka University, Machikaneyama 1-1, Toyonaka, Osaka 560-0043}
\email{bessho@vega.ess.sci.osaka-u.ac.jp}

%


\maketitle

\begin{abstract}
Collapse and fragmentation of primordial filamentary clouds under isotropic dissociation radiation is investigated with one-dimensional hydrodynamical calculations. We investigate the effect of dissociation photon on the filamentary clouds with calculating non-equilibrium chemical reactions. With the external radiation assumed to turn on when the filamentary cloud forms, the filamentary cloud with low initial density ($n_0 \le 10^2 \mathrm{cm^{-3}}$) suffers photodissociation of hydrogen molecules. In such a case, since main coolant is lost, temperature increases adiabatically enough to suppress collapse. As a result, the filamentary cloud fragments into very massive clouds ($\sim 10^5 M_\odot$). On the other hand, the evolution of the filamentary clouds with high initial density ($n_0>10^2 \mathrm{cm^{-3}}$) is hardly affected by the external radiation. This is because the filamentary cloud with high initial density shields itself from the external radiation. It is found that the external radiation increases fragment mass. This result is consistent with previous results with one-zone models. It is also found that fragment mass decreases owing to the external dissociation radiation in the case with sufficiently large line mass.
\end{abstract}

\section{Introduction}
It is accepted that the density perturbations collapse and cool due to hydrogen molecules ($\mathrm{H_2}$) to form so called population III (popIII) stars (Bromm et al. 1999, 2002; Abel et al. 2000, 2002; Yoshida et al. 2008). PopIII is expected to form in halos with $\ge 10^6 M_\odot$ (Tegmark et al. 1997). If PopIII is massive star, it is expected to affect neighbor clouds via radiative feedbacks. Radiative feedbacks cause ionization and dissociation. Although ionization photon tends to be prevented from spreading out of halos because of large opacity of hydrogen atoms, dissociation photon tends to spread out of halos (Kitayama et al. 2004). Thus, some regions are expected not to be ionized but to be photodissociated. We consider the filamentary clouds in such a region.

Filamentary clouds are possible origin of stars. In general, non-spherical gas cloud tends to become sheet-like cloud, and sheet-like cloud tends to fragment into the filamentary clouds (Miyama et al. 1987). In numerical cosmological simulations of first star formation, filamentary structure is frequently seen (e.g., Abel et al. 1998; Bromm et al. 1999; Greif et al. 2006). Recently, many filamentary structures have been found through the $Herschel$ Gould Belt Survey (Andr\'e et al. 2010). A filamentary cloud is possible to fragment into many quasi-spherical clouds (Nagasawa 1987; Inutsuka $\&$ Miyama 1997). These spherical clouds are expected to become stars or other astronomical objects. We investigate how this process exceeds when first stars form.

There are previous works about fragmentation of the filamentary cloud (Uehara et al. 1996 ; Nakamura $\&$ Umemura 1999, 2001, 2002; Flower 2002; Omukai $\&$ Yoshii 2003). Among these, Nakamura $\&$ Umemura (1999, 2001, 2002) used one-dimensional hydrodynamical calculations and two-dimensional hydrodynamical calculations. The authors considered many cases with various initial density, temperature, line mass, and initial fraction of $\mathrm{H_2}$. It is found that fragment mass is $1-500M_\odot$ and has a bimodal distribution when initial $\mathrm{H_2}$ fraction is $10^{-3}$. However, the effect of the external radiation was not considered. Hence, these works is applicable only to first star formation. Among works mentioned above, Omukai $\&$ Yoshii (2003) considered the external dissociation radiation. Their work is applicable to formation of second-generation stars. They calculated the thermal evolution of the filamentary cloud under the isotropic external radiation with one-zone model assuming free-fall. They assumed that the filamentary cloud fragments when its density becomes $100$ times higher than the loitering point. Under this assumption, Omukai $\&$ Yoshii (2003) concluded that the effect of the external dissociation radiation decreases fragment mass.

In Bessho $\&$ Tsuribe (2012) (hereafter Paper I), we investigated collapse and fragmentation of a filamentary clouds under the isotropic external radiation using one-zone models. We assumed that the external radiation turns on when the filamentary cloud forms. By taking into accounts pressure effect explicitly, we found that the filamentary clouds with low initial density ($n_0 \le 10^2 \mathrm{cm^{-3}}$) suffer photodissociation and fragment into very massive clouds ($\sim 10^{4-5} M_\odot$). It was found that the effect of the external radiation increases fragment mass. The evolution of the filamentary clouds with high initial density ($n_0 > 10^2 \mathrm{cm^{-3}}$) or with sufficiently large line mass is not affected by the external radiation owing to self-shielding. In Paper I, at first, we assumed the uniform filamentary clouds with homologous collapse. However, in realistic situations, filamentary clouds are expected to collapse in run-away fashion. Hence, in Paper I, we also introduced "rarefied filament model" as an improved model which includes the effect of run-away collapse partly. In rarefied filament model, line mass of collapsing core decreases as rarefaction wave propagates from the cloud surface. As a result of the rarefied filament model, we found that the effect of the external radiation increases fragment mass. However, this result is apparently inconsistent with Omukai $\&$ Yoshii (2003).

The purpose of the present paper is to investigate whether or not fragment mass increases owing to the effect of the external radiation and to clarify the reason why our result is inconsistent with Omukai $\&$ Yoshii (2003). For this purpose, we extend our previous investigation using the one-dimensional model which includes the full characteristics of run-away collapse and have more realistic results. Furthermore, in this paper, we also consider the further evolution of each fragment.

We assume that the external radiation is isotropic. Intensity of the external dissociation radiation is set according to distribution of intensity of the dissociation radiation at redshift $z \sim 10$ (Dijkstra et al. 2008). The intensity which we consider is moderate ($\S 2.2$), and we consider the situation where the dissociation radiation originates from halos out of a halo including the filamentary cloud.

In Paper I, we concentrated on the case where the external radiation turns on when the filamentary cloud forms ($n \ge 10\mathrm{cm^{-3}}$). To clarify the reason of the apparent difference of the conclusions between us and Omukai $\&$ Yoshii (2003), we also consider the case where the external radiation turns on at lower density ($n=0.1\mathrm{cm^{-3}}$) as in Omukai $\&$ Yoshii (2003).

In $\S 2$, we describe our model for the filamentary clouds. We present numerical results in $\S 3$. In $\S 4$, we investigate the reason for the difference between us and Omukai $\&$ Yoshii (2003). $\S 5$ is devoted to conclusions and discussion.

\section{Model for the filamentary clouds}
\subsection{Basic equations}
We assume the axisymmetric filamentary cloud. We do not consider dark matter for simplicity. This simplicity gives us a good approximation in the case with high initial density. In the case with low initial density, the effect of dark matter gravity is underestimated. Hydrodynamical equation of motion in Lagrangian form is given by
\begin{eqnarray}
\frac{Dv}{Dt}=-\frac{2Gl}{r}-2\pi r \frac{\partial P}{\partial l},
\end{eqnarray}
where $v$ is velocity in the cylindrical radial direction, $G$ is gravitational constant, $l$ is line mass within cylindrical radius $r$, and $P$ is pressure for ideal gas given by
\begin{eqnarray}
P=n k_B T,
\end{eqnarray}
with number density $n$, temperature $T$, and Boltsmann constant $k_B$.

We also solve energy equation given by
\begin{eqnarray}
\frac{du}{dt}=-P\frac{d\ }{dt}\frac{1}{\rho}-\frac{\Lambda_{\mathrm{net}}}{\rho}
\end{eqnarray}
where $\rho$ is density and $u$ is thermal energy per unit mass,
\begin{eqnarray}
u=\frac{1}{\gamma_{\mathrm{ad}}-1} \frac{k_B T}{\mu m_{\mathrm{H}}}
\end{eqnarray}
with adiabatic index $\gamma_{\mathrm{ad}}$, mean molecular weight $\mu$, and mass of a hydrogen atom $m_{\mathrm{H}}$. The symbol $\Lambda_{\mathrm{net}}$ in equation (3) is cooling rate per unit volume including lines of $\mathrm{H}$, lines of $\mathrm{H_2}$, lines of $\mathrm{HD}$, and chemical heating/cooling. Since continuum processes hardly change the evolution of the filamentary clouds (Paper I), we neglect them. As for lines, we estimate cooling rate from detailed balance of population of energy levels. The escape probability for emission by the transition between level $i$ and $j$ is given by
\begin{eqnarray}
\beta_{ij}=\frac{1-e^{-\tau_{ij}}}{\tau_{ij}},
\end{eqnarray}
with assuming the velocity profile $v_r (r) \propto r$ (Castor 1970). The cooling rate is multiplied by this escape probability. Optical depth, $\tau_{ij}$, is given by
\begin{eqnarray}
\tau_{ij}&=&\int^{R_{\mathrm{out}}}_r \kappa_{ij} (r^\prime) \ dr^\prime \nonumber \\
&=&\int^{R_{\mathrm{out}}}_r \frac{h \nu_{ij}}{8 \pi \Delta \nu_{ij}} (n_j B_{ji}-n_i B_{ij}) dr^\prime,
\end{eqnarray}
where $R_{\mathrm{out}}$ is radius of the outer boundary, $\kappa_{ij}$ is opacity for lines, $h \nu_{ij}$ is energy difference between levels $i$ and $j$, $n_i$ ($n_j$) is the level population at level $i$ ($j$), $B_{ij}$ and $B_{ji}$ are the Einstein $B$-coefficients, and $\Delta \nu_{ij}=\nu_{ij}/c \sqrt{2k_B T/\mu m_{\mathrm{H}}}$ is the thermal Doppler width of transition line $i \rightarrow j$.

We consider non-equilibrium chemical reactions by solving following equation for each fluid element :
\begin{eqnarray}
\frac{df_i}{dt}=\sum_{j,\ k} k_{ijk} f_j f_k n + \sum_j k_{ij} f_j ,
\end{eqnarray}
where $k_{ijk}$ and $k_{ij}$ are reaction rates for formation and destruction of speices $i$, and $f_i$ is fraction of species $i$. We consider fourteen species : $\mathrm{H}$, $\mathrm{H^+}$, $\mathrm{H^-}$, $\mathrm{H_2}$, $\mathrm{H_2^+}$, $\mathrm{He}$, $\mathrm{He^+}$, $\mathrm{He^{++}}$, $\mathrm{D}$, $\mathrm{D^+}$, $\mathrm{D^-}$, $\mathrm{HD}$, $\mathrm{HD^+}$, and $\mathrm{e^-}$. We consider 26 chemical reactions concerned with $\mathrm{H}$ and $\mathrm{He}$ taken from Nakamura $\&$ Umemura (2001), photodissociation of $\mathrm{H_2}$ (equation 10), and 18 chemical reactions concerned with $\mathrm{H}$ and $\mathrm{D}$ taken from Nakamura $\&$ Umemura (2002). We solve equation (7) with implicit integrator.

We solve equations (1)-(7) using 200 spatial meshes in cylindrical radial direction. As for initial interval of meshes, we set $\Delta r_{i+1} = 1.01 \Delta r_i$, where $\Delta r_i \equiv r_{i+1}-r_i$. Hence, spatial resolution is much better in the central region. The mesh size $\Delta r_i$ is checked to be shorter than $1/4$ of local Jeans length\footnote{The exact definition of Jeans length is given by
\begin{eqnarray}
\lambda_{\mathrm{J}}=\biggl(\frac{\pi}{G \mu m_{\mathrm{H}} n_c} \biggr)^{1/2} c_s.
\end{eqnarray}
} at all times (Truelove et al. 1997). We set outer boundary to be $10R_0$ as a sufficiently large value, where $R_0$ is the effective radius given by
\begin{eqnarray}
R_0=\sqrt{\frac{2fk_BT_0}{\pi G \mu^2 m_{\mathrm{H}}^2 n_0}},
\end{eqnarray}
with line mass parameter $f$ (equation 17), initial temperature $T_0$, and initial number density at the center $n_0$. As for the outer boundary condition, we assume that the external pressure is zero.

\begin{figure}
\begin{center}
\begin{tabular}{cc}
\resizebox{80mm}{!}{\includegraphics{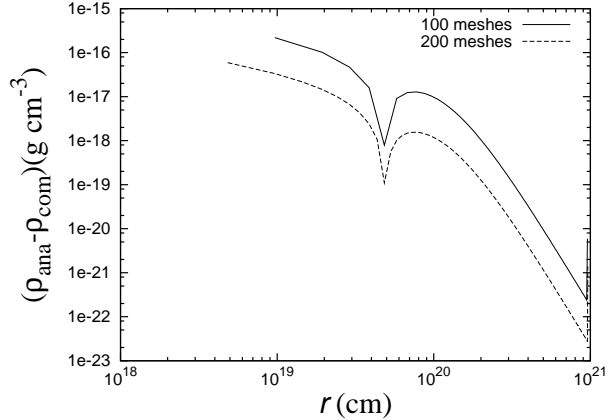}} \\
\end{tabular}
\caption{Error of density distribution of isothermal equilibrium filamentary cloud ($T=300\mathrm{K}$) for the numerical results with $100$ meshes and $200$ meshes. The symbol $\rho_{\mathrm{ana}}$ is analytic solution, and $\rho_{\mathrm{com}}$ is numerical results.}
\end{center}
\end{figure}

We solve equation (1) in the second-order-accurate finite-difference scheme with the artificial viscosity (Richtmyer $\&$ Morton 1967). We calculate in the same way as Thoul $\&$ Weinberg (1995), except for cylindrical geometry. To check accuracy of our code, first, we calculate the density distribution of isothermal equilibrium filamentary cloud ($T=300\mathrm{K}$) with 100 and 200 meshes, and fixed time step. Drag term $-2v(r)/dt$ is added to equation of motion and is eliminated after the step number reaches $125$ for case with $100$ meshes and $250$ for case with $200$ meshes. After $250$ steps ($100$ meshes) and $500$ steps ($200$ meshes), we have error of density distribution shown in figure 1. Error for case with $100$ meshes is $4$ times of error for case with $200$ meshes. Hence, our code is second-order-accurate in space.

Second, we check temporal accuracy of our code in time. We calculate collapse of pressure-less uniform filamentary cloud in free-fall state. Analytic solution is given by
\begin{eqnarray}
\int^{\sqrt{-\log F(t)}}_{0} e^{-x^2} dx = \sqrt{\pi G \rho(0)} t,
\end{eqnarray}
where $F(t)=r(t)/r(0)$, $\rho(0)$ is the initial density. Time step is set to be $\Delta t=10^{-4} t_{\mathrm{ff}}(t=0)$ and $2\Delta t$. Calculation is continued until the density becomes $100$ times of the initial density. In the case with $\Delta t$, $F$ at the end of calculation is $0.0913290$. In the case with $2\Delta t$, $F$ at the end of calculation is $0.09132307$. Analytic solution predicts $F=0.9132284$. Error with $2\Delta t$ is $4$ times larger than $\Delta t$. Hence, our code has second-order-accuracy in time.

\subsection{External radiation}
We assume that the external dissociation radiation is isotropic. Intensity of the external radiation is set according to Dijkstra et al. (2008). Dijkstra et al. (2008) calculated the probability distribution of mean intensity of the dissociation radiation at redshift $z \sim 10$ by estimating mean intensity of the dissociation photon from the surrounding halos to a single halo. In this paper, we use mean intensities whose probability is $0.4$ and $0.06$ as in Paper I. The external radiation is assumed to be thermal radiation of $120M_\odot$ star, and we determine surface temperature ($T_{\mathrm{sur}}=95719\mathrm{K}$) according to Schaerer (2002). We assume that the external radiation turns on when the filamentary clouds forms in $\S 3$. In $\S 4$, we change the density when the external radiation turns on.

We calculate the photodissociation reaction of $\mathrm{H_2}$,
\begin{eqnarray}
\mathrm{H_2} + \gamma \rightarrow \mathrm{H_2}^\ast \rightarrow 2 \mathrm{H},
\end{eqnarray}
(solomon process) where $\gamma$ is photon with $12.4 \mathrm{eV}$ and $\mathrm{H_2}^\ast$ is excited state of $\mathrm{H_2}$. The reaction rate is given by
\begin{eqnarray}
k_{\mathrm{2step}} = 1.4 \times 10^9 f_{\mathrm{sh}} J_{\mathrm{ext}} \ \mathrm{s^{-1}},
\end{eqnarray}
where $J_{\mathrm{ext}}$ is mean intensity of the external radiation at the surface of the filamentary cloud and $f_{\mathrm{sh}}$ is self-shielding function,
\begin{eqnarray}
f_{\mathrm{sh}}=\mathrm{min} \biggl[ 1,\ \biggl(\frac{N_{\mathrm{H_2}}}{10^{14} \mathrm{cm^{-2}}} \biggr)^{-3/4} \biggr],
\end{eqnarray}
where $N_{\mathrm{H_2}}$ is column density of $\mathrm{H_2}$ (Draine $\&$ Bertoldi 1996) estimated as
\begin{eqnarray}
N_{\mathrm{H_2}}(r)= \int^{R_{\mathrm{out}}}_{r} n_{\mathrm{H_2}} (r^\prime) dr^\prime.
\end{eqnarray}

\subsection{Fragmentation of the filamentary cloud}
There are two important timescales during the collapse of filamentary clouds. One is timescale for density evolution, $t_{\mathrm{dyn}} \equiv \rho_c/\dot{\rho_c}$, and the other is timescale for fragmentation, $t_{\mathrm{frag}} \equiv 5.17/\sqrt{2\pi \rho_c G}$ (Nagasawa 1987). The latter is the timescale during which the fastest growing mode of perturbation grows to non-linear. If the fastest growing mode has time enough to grow to non-linear before the fastest growing mode changes, the filamentary cloud is expected to fragment. Thus, we assume that the filamentary clouds fragment when $t_{\mathrm{dyn}}>t_{\mathrm{frag}}$. Once condition for fragmentation is satisfied, it has been satisfied after that.

We estimate fragment mass by integrating region with the density higher than 10$\%$ of central density $n_c$,
\begin{eqnarray}
M_{\mathrm{frag}}=\lambda_{\mathrm{frag}} \int^{r(n=0.1n_c)}_{0} 2\pi r^\prime \rho dr^\prime,
\end{eqnarray}
where
\begin{eqnarray}
\lambda_{\mathrm{frag}}=2\pi \frac{c_s}{0.288 \sqrt{4 \pi G \mu m_{\mathrm{H}} n_c}}
\end{eqnarray}
is the wave length of the fastest growing mode for equilibrium filamentary clouds (Nagasawa 1987) and $c_s$ is sound speed. Since $\lambda_{\mathrm{frag}} \propto \rho^{-1/2}$, fragment mass is smaller when the filamentary cloud reaches higher density before fragmentation. Since interval of integration in equation (14) is approximately Jeans length, Jeans mass at fragmentation is close to fragment mass (see $\S 3.3$).

\subsection{Parameters and initial conditions}
In this paper, we treat three physical quantities as parameters. First is initial density $n_0$, second is normalized mean intensity of the external radiation,
\begin{eqnarray}
J_{21} \equiv \frac{J_{h\nu=13.6\mathrm{eV},\mathrm{ext}}}{10^{-21} \mathrm{erg\ cm^{-2} s^{-1} Hz^{-1} sr^{-1}}},
\end{eqnarray}
and third is line mass parameter\footnote{If a filamentary cloud forms as a result of fragmentation of the fastest growing mode in a sheet-like cloud, the typical value of $f$ is $2$ (Miyama et al. 1987).},
\begin{eqnarray}
f \equiv \frac{\pi G \mu^2 m_{\mathrm{H}}^2 n_0}{2k_B T_0} R_0^2.
\end{eqnarray}
Line mass parameter is important in the view point of dynamical evolution. Initial density $n_0$ is important for thermal evolution. With respect to photodissociation, mean intensity $J_{21}$ and initial density $n_0$ are important.

We consider cases with $\log_{10} n_0=1$, $1.5$, $2$, $2.5$, $3$, $3.5$, $4$, $4.5$, $5$, $5.5$, and $6$ for $n_0$ and $f=1.5$, $2$, $2.5$, $3$, $3.5$, $4$, $5.5$, and $6$ for $f$. For $J_{21}$, we consider $J_{21}=0$, $1$, $6.5$, and $10$. According to Dijkstra et al. (2008), the case with $J_{21}=1$ represents the weak external radiation case, $J_{21}=6.5$ is mean intensity with the highest probability ($0.4$), and $J_{21}=10$ represents the strong radiation case whose probability is $0.06$.

We assume initial density distribution as
\begin{eqnarray}
n(r)=n_0 \biggl( 1+ \frac{r^2}{R_0^2} \biggr)^{-2}
\end{eqnarray}
with $R_0$ given in equation (8). With $f=1$, equation (18) represents equilibrium density distribution for the isothermal filamentary cloud (Ostriker 1964). In this paper, we concentrate on collapsing filamentary cloud with $f>1$. Initial velocity distribution is assumed to be
\begin{eqnarray}
v(r)=- \frac{c_s}{R_0+ \sqrt{R_0^2+r^2}} r.
\end{eqnarray}
Equation (19) indicates that infall velocity is proportion to radius at $r \rightarrow 0$ and is constant ($\sim -c_s$) at $r \rightarrow \infty$. Although the actual initial velocity may be different from equation (19) depending on the detail of dynamical evolution of the filamentary cloud formation, results in this paper will not change qualitatively unless initial velocity is much faster than a few times of equation (19).

As initial temperature, we adopt $T_0=300\mathrm{K}$, assuming that the filamentary cloud forms after the cloud undergoes $\mathrm{H_2}$ cooling. We also adopt $f_{\mathrm{H_2}}=10^{-4}$ and $f_{\mathrm{e}}=10^{-4}$. As for the value of $f_{\mathrm{H_2}}$, we refer the result in Paper I. Fraction of electron is set in order not to change $f_{\mathrm{H_2}}$ artificially via $\mathrm{H^-}$ channel. Initial fraction of proton is set to be $f_{\mathrm{p}}=10^{-4}$ for charge conservation. We assume $[\mathrm{H}]/[\mathrm{D}]=4 \times 10^{-5}$, which is consistent with observations of the deuterium Ly$\alpha$ feature (e.g., O'Meara et al. 2001). Initial fraction of the others is set to be zero.

\section{Results of one-dimensional hydrodynamical calculations}
\subsection{Cases without the external radiation}
To investigate the effect of the external radiation, at first, we show the results of the cases without the external radiation.

\subsubsection{Low density filamentary cloud with small line mass}
\begin{figure}
\begin{center}
\begin{tabular}{cc}
\resizebox{80mm}{!}{\includegraphics{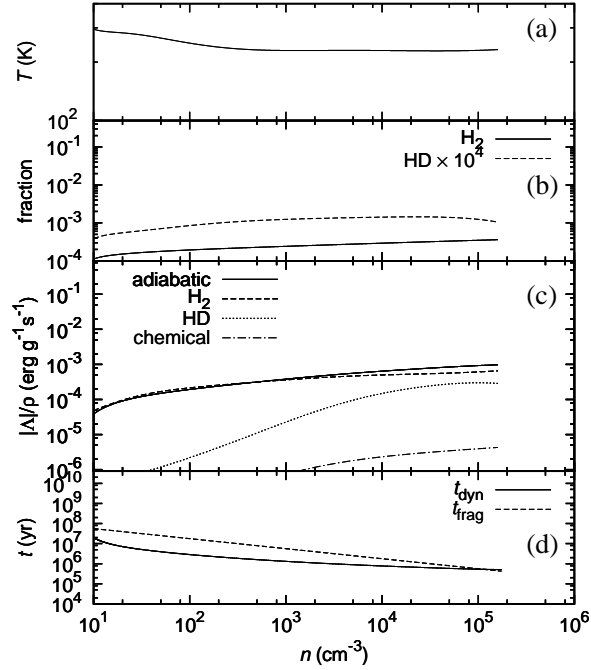}} \\
\end{tabular}
\caption{Evolution of the temperature (a), $f_{\mathrm{H_2}}$ and $f_{\mathrm{HD}} \times 10^4$ (b), the heating and cooling rate (c), and $t_{\mathrm{dyn}}$ and $t_{\mathrm{frag}}$ (d), respectively, as a function of the central density for the case with $(f,n_0,J_{21})=(1.5,10 \mathrm{cm^{-3}},0)$. In the diagram (c), "adiabatic" denotes the adiabatic heating, "$\mathrm{H_2}$" does the $\mathrm{H_2}$ line cooling, "$\mathrm{HD}$" does the $\mathrm{HD}$ line cooling, and "chemical" does the chemical heating or cooling.}
\end{center}
\end{figure}

First, we show the results for the case with low initial density and small line mass, $(f,n_0,J_{21})=(1.5,10\mathrm{cm^{-3}},0)$ (figure 2). In the early stage of collapse, $\mathrm{H_2}$ cooling dominates adiabatic heating a little, and temperature decreases. After the density reaches $n \sim 10^3 \mathrm{cm^{-3}}$, adiabatic heating dominates $\mathrm{H_2}$ cooling. In $n > n_{\mathrm{crit}} \sim 10^4 \mathrm{cm^{-3}}$, $\mathrm{H_2}$ cooling rate is proportional to $n$, while it is proportional to $n^2$ in $n < n_{\mathrm{crit}}$. Hence, since cooling time becomes constant in $n > n_{\mathrm{crit}}$, dynamical time becomes constant and longer than fragmentation time. When $n$ reaches $\sim 2 \times 10^5 \mathrm{cm^{-3}}$, condition for fragmentation is satisfied with $M_{\mathrm{frag}} \sim 1220 M_\odot$. To ensure that fragmentation occurs, we continue to calculate the evolution until free-fall time has past after the condition for fragmentation is first satisfied. Once condition for fragmentation is satisfied, it has been satisfied.

\subsubsection{Low density filamentary cloud with large line mass}
\begin{figure}
\begin{center}
\begin{tabular}{cc}
\resizebox{80mm}{!}{\includegraphics{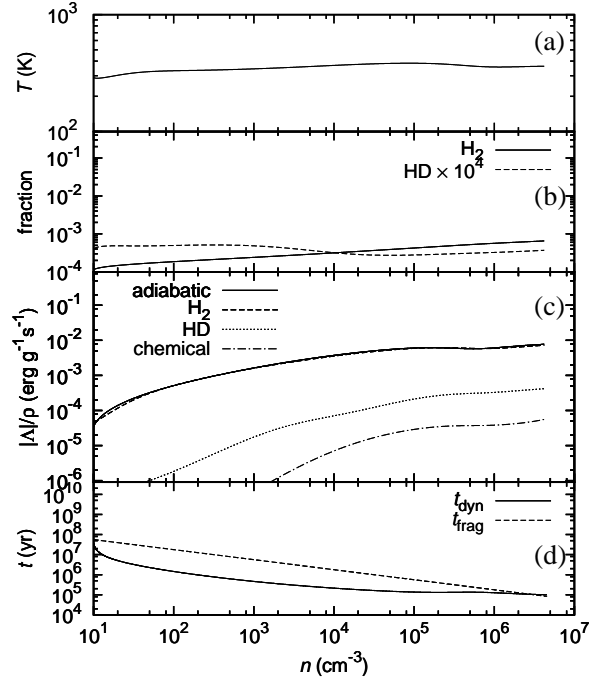}} \\
\end{tabular}
\caption{Same as figure 2, but $(f,n_0,J_{21})=(6,10\mathrm{cm^{-3}},0)$.}
\end{center}
\end{figure}

Next, we show the result for the case with low initial density and large line mass, $(f,n_0,J_{21})=(6,10\mathrm{cm^{-3}},0)$ (figure 3). Owing to larger line mass, the filamentary cloud collapses to higher density than figure 2. Collapse continues up to high density ($n \sim 10^6 \mathrm{cm^{-3}}$), and the filamentary cloud fragments. Fragment mass is $\sim 490 M_\odot$. Until fragmentation, adiabatic heating and $\mathrm{H_2}$ cooling balance, and temperature is approximately constant ($T \sim 350 \mathrm{K}$).

\subsubsection{High density filamentary cloud with small line mass}
\begin{figure}
\begin{center}
\begin{tabular}{cc}
\resizebox{80mm}{!}{\includegraphics{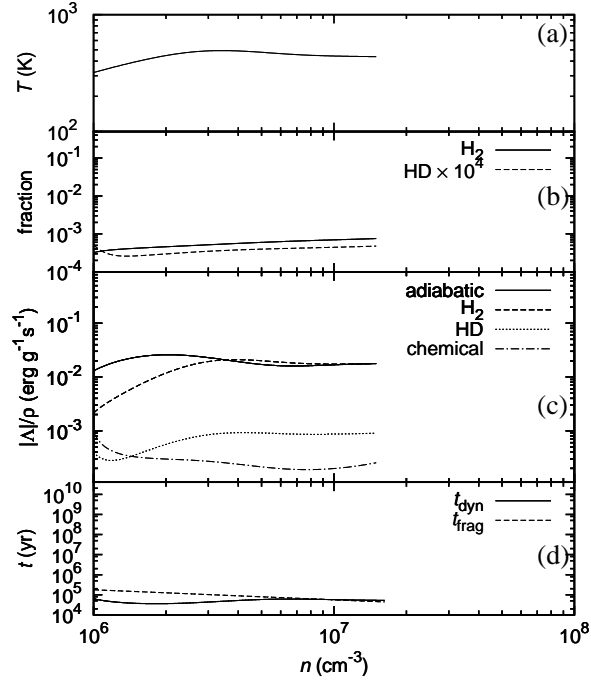}} \\
\end{tabular}
\caption{Same as figure 2, but $(f,n_0,J_{21})=(1.5,10^6\mathrm{cm^{-3}},0)$.}
\end{center}
\end{figure}

As the final example, we show the result for the case with high initial density and small line mass, $(f,n_0,J_{21})=(1.5,10^6\mathrm{cm^{-3}},0)$ (figure 4). In the early stage of collapse, Adiabatic heating dominates $\mathrm{H_2}$ cooling and temperature increases. During collapse, $\mathrm{H_2}$ cooling rate increases and approximately balances with adiabatic heating rate at $n \sim 3 \times 10^6 \mathrm{cm^{-3}}$. After temperature decreases a little, fragmentation condition is satisfied at $n \sim 10^7 \mathrm{cm^{-3}}$ since collapse is suppressed owing to high temperature. Fragment mass is $\sim 370 M_\odot$.

In summary, in the cases without the external radiation, the filamentary cloud undergoes approximately isothermal states and fragments. This feature comes from the fact that $\mathrm{H_2}$ cooling and adiabatic heating compete each other. The results in this subsection are similar to previous works (Nakamura $\&$ Umemura 2001, 2002). For the parameter set same as figure 2, in Paper I, fragment mass was $23 M_\odot$ in the uniform model and $3500 M_\odot$ in rarefied filament model. Fragment mass $1220 M_\odot$ in the one-dimensional model is close to the results of the rarefied filament model. This result indicates that the effect of run-away collapse is important to estimate fragment mass.

\subsection{Cases with the external radiation}
In this subsection, using the same parameters as figure 2, 3, and 4, we investigate how the external radiation changes the thermal evolution and fragment mass of the filamentary cloud.

\subsubsection{Low density filamentary cloud with small line mass and strong radiation}
\begin{figure}
\begin{center}
\begin{tabular}{cc}
\resizebox{80mm}{!}{\includegraphics{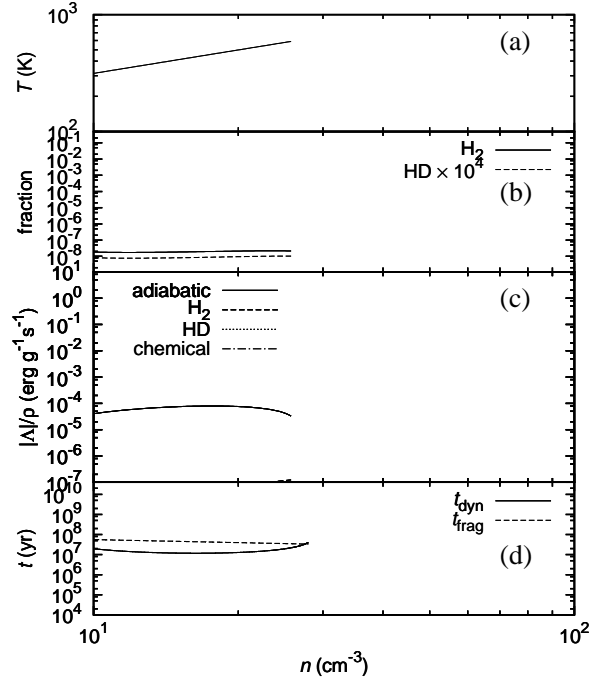}} \\
\end{tabular}
\caption{Same as figure 2, but $(f,n_0,J_{21})=(1.5,10\mathrm{cm^{-3}},10)$.}
\end{center}
\end{figure}

First, we show the result for the case with low initial density, small line mass, and strong external radiation, $(f,n_0,J_{21})=(1.5,10\mathrm{cm^{-3}},10)$ (figure 5), where the external radiation is added to the case of figure 2. This case would be affected by the external radiation because of low density. Most of $\mathrm{H_2}$ are photodissociated in the early stage of collapse, and temperature increases adiabatically. The filamentary cloud fragments into very massive fragments ($\sim 2.4 \times 10^5 M_\odot$) at $n \sim 30 \mathrm{cm^{-3}}$. This result demonstrates that in the case with low initial density and small line mass the external radiation increases fragment mass and changes the thermal evolution.

\subsubsection{Low density filamentary cloud with large line mass and strong radiation}
\begin{figure}
\begin{center}
\begin{tabular}{cc}
\resizebox{80mm}{!}{\includegraphics{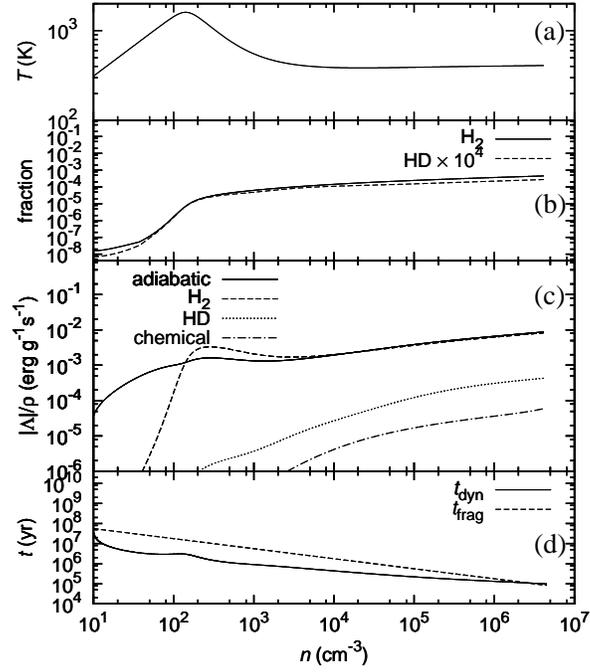}} \\
\end{tabular}
\caption{Same as figure 2, but $(f,n_0,J_{21})=(6,10\mathrm{cm^{-3}},10)$.}
\end{center}
\end{figure}

Next, we show the result for the case with low initial density, large line mass, and strong external radiation, $(f,n_0,J_{21})=(6,10\mathrm{cm^{-3}},10)$ (figure 6), where the external radiation is added to the case of figure 3. This case is also excepted to be affected by the external radiation because of low density. However, the filamentary cloud may collapse up to higher density than figure 5 because of large line mass. In figure 6, it is seen that most of $\mathrm{H_2}$ is photodissociated in the early stage of collapse and temperature increases adiabatically as in figure 5. However, since the filamentary cloud is more massive than in figure 5, stronger gravity and large inertia help collapse. Fragmentation does not occur during early adiabatic phase, and collapse continues until the density becomes higher than in figure 5. At $n \sim 10^2 \mathrm{cm^{-3}}$, $\mathrm{H_2}$ starts to form and shields itself from the external radiation. The filamentary cloud starts to cool owing to $\mathrm{H_2}$ cooling. After $n \sim 10^3 \mathrm{cm^{-3}}$, Since $\mathrm{H_2}$ cooling balances with adiabatic heating, temperature becomes nearly constant ($T \sim 400 \mathrm{K}$). The filamentary cloud fragments into clouds with $\sim 590 M_\odot$ at $n \sim 10^6 \mathrm{cm^{-3}}$. Fragment mass and density at fragmentation are similar to the case without the external radiation (figure 3). In the case with large line mass ($f=6$), it is found that fragment mass is hardly affected by the external radiation although the evolution of temperature is affected by the external radiation in the early stage of collapse.

\subsubsection{High density filamentary cloud with small line mass and strong radiation}
\begin{figure}
\begin{center}
\begin{tabular}{cc}
\resizebox{80mm}{!}{\includegraphics{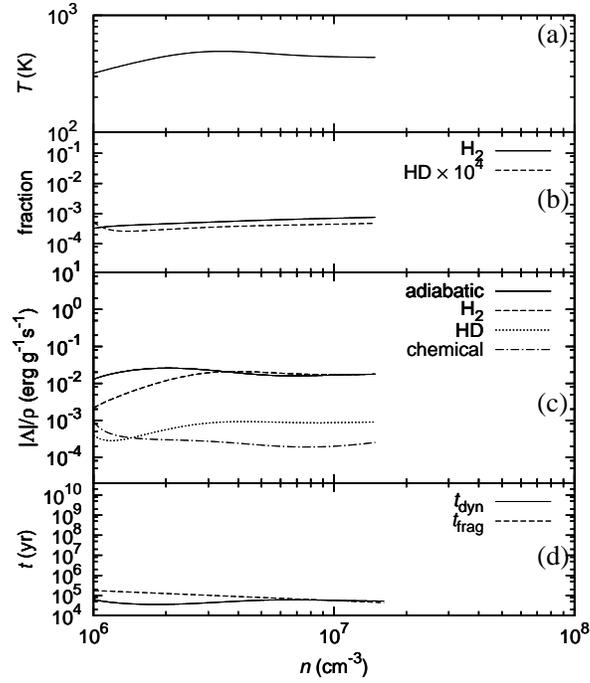}} \\
\end{tabular}
\caption{Same as figure 2, but $(f,n_0,J_{21})=(1.5,10^6\mathrm{cm^{-3}},10)$.}
\end{center}
\end{figure}

Finally, we show the result for the case with high initial density, small line mass, and strong external radiation, $(f,n_0,J_{21})=(1.5,10^6\mathrm{cm^{-3}},10)$ (figure 7), where the external radiation is added to the case of figure 4. This case may not be affected by the external radiation because of high density. Since initial density is high enough to shield the filamentary cloud from the external radiation, $\mathrm{H_2}$ near the center of cloud is not photodissociated. The evolution of temperature at the center is hardly affected by the external radiation and is similar to figure 4. The filamentary cloud fragments into clouds with $\sim 400 M_\odot$ at $n \sim 10^7 \mathrm{cm^{-3}}$. It is found that the effect of the external radiation is not important in the case with high initial density ($n_0=10^6 \mathrm{cm^{-3}}$).

In summary, in the case with low initial density ($n_0<10^2\mathrm{cm^{-3}}$), the filamentary cloud suffers photodissociation in the early stage of collapse. In such case, temperature increases adiabatically. The filamentary cloud with small line mass ($f=1.5$) fragments during adiabatic phase. On the other hand, the filamentary cloud with large line mass ($f=6$) does not fragment during adiabatic phase and collapses with shielding itself from the external radiation. In this case, fragment mass is hardly affected by the external radiation. In the case with high initial density ($n_0=10^6 \mathrm{cm^{-3}}$), the thermal evolution of the filamentary cloud is hardly affected by the external dissociation radiation.

One-zone model predicts fragment mass different from $2.4 \times 10^5 M_\odot$ which one-dimensional model predicts. For example, for the parameter set same as figure 5, in Paper I, the uniform model predicted $1.5 \times 10^5 M_\odot$ and rarefied model predicted $2.7 \times 10^4 M_\odot$. Difference between one-zone models and one-dimensional model originates from difference of dynamical equation (virial equation in one-zone model and hydrodynamical equation of motion in one-dimensional model). In one-dimensional model, collapse is run-away collapse, and fragmentation condition is satisfied at lower density since free-fall time balances with sound crossing time in the central dense region. Hence, in one-dimensional model, fragment mass becomes larger than in one-zone model.

\subsection{Property of the filamentary cloud at fragmentation}
In this subsection, we show the profile of physical quantities (density, temperature, in-fall velocity, and ratio of pressure gradient to gravitational force) at fragmentation. We focus on density profile and investigate whether or not the universal profile at fragmentation exists. Moreover, we compare fragment mass (equation 14) with Jeans mass estimated with the central density and temperature.

\subsubsection{Case without the external radiation}
\begin{figure}
\begin{center}
\begin{tabular}{cc}
\resizebox{80mm}{!}{\includegraphics{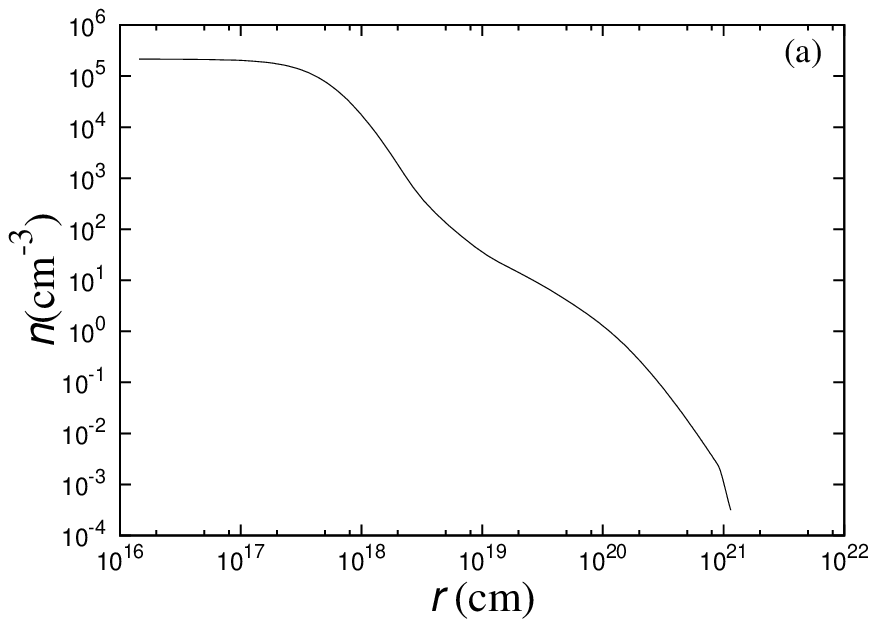}} &
\resizebox{80mm}{!}{\includegraphics{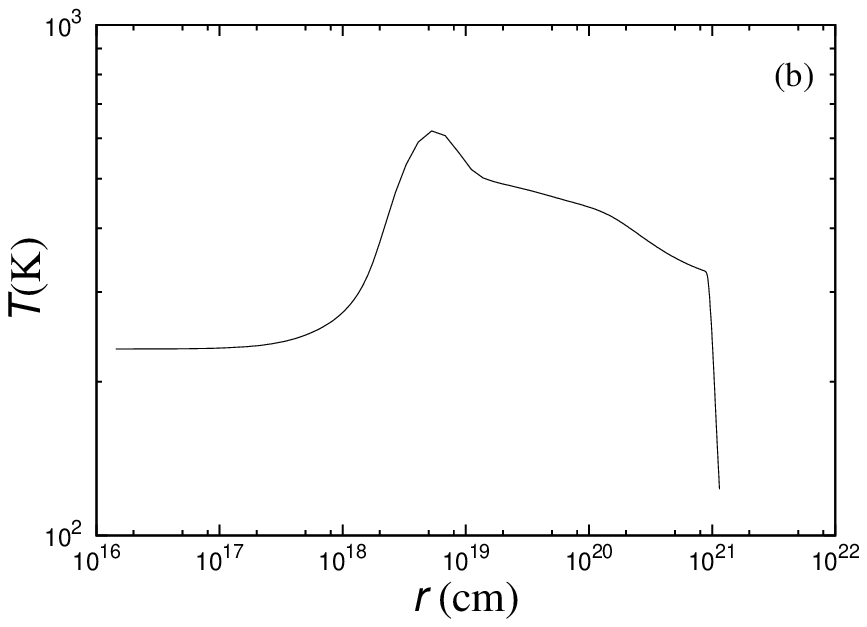}} \\
\resizebox{80mm}{!}{\includegraphics{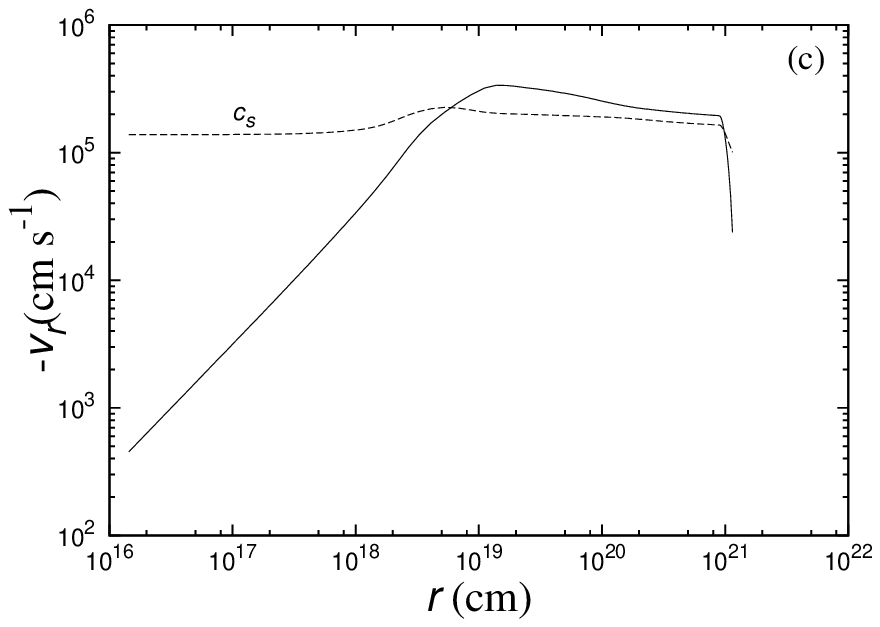}} &
\resizebox{80mm}{!}{\includegraphics{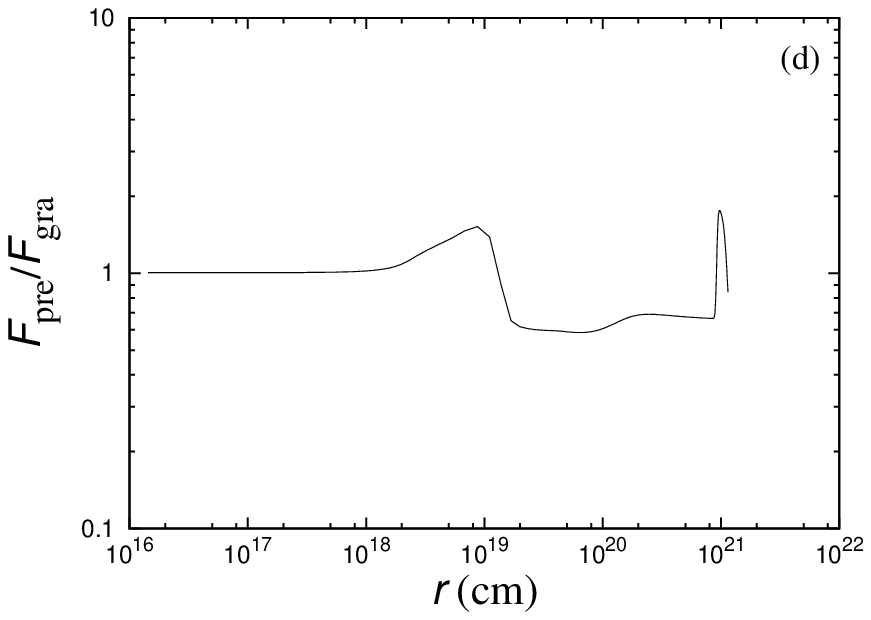}} \\
\end{tabular}
\caption{Profile of density (a), temperature (b), in-fall velocity profile (c), and ratio of pressure gradient to gravity (d) at fragmentation of the filamentary cloud with $(f,n_0,J_{21})=(1.5,10\mathrm{cm^{-3}},0)$.}
\end{center}
\end{figure}

We show the profiles of density, temperature, infall velocity, and ratio of pressure gradient to gravitational force at fragmentation in the case with $(f,n_0,J_{21})=(1.5,10\mathrm{cm^{-3}},0)$ (figure 8). In diagram (a) in figure 8, it is seen that dense central region within Jeans length ($\lambda_J \sim 2.3 \times 10^{18} \mathrm{cm}$) has uniform density, and the density profile in the outer envelope is proportional to $r^{-4}$. This density profile is similar to that of equilibrium solution for the isothermal filamentary cloud (Ostriker 1967). However, between $r=10^{19}\mathrm{cm}$ and $r=10^{20}\mathrm{cm}$, slope of the density profile is sallower than $r^{-4}$. Temperature is highest outside $r_{\mathrm{cool}} \sim 2 \times 10^{18} \mathrm{cm}$ where $t_{\mathrm{cool}}=t_{\mathrm{ff}}$, and pressure gradient force is stronger than gravity force outside $r_{\mathrm{cool}}$. Hence, matter is pushed outward. Velocity profile is in proportion to radius in the central dense region and is constant larger than sound speed in the outer envelope. Ratio of pressure gradient to gravity is nearly $1$ ($\sim 1.01$) inside $r_{\mathrm{cool}}$. In diagram (b), drop of temperature at the surface is seen. Since we assume that the external pressure is zero, adiabatic cooling occurs at several meshes of the surface. Moreover, these meshes are pushed by inner meshes with higher pressure and fall more slowly than inner meshes. However, these effects do not affect the central region. Fragment mass is $1220M_\odot$ which is close to Jeans mass ($1140M_\odot$) estimated with central density and temperature. Since $t_{\mathrm{dyn}}$ is about 6 times of $t_{\mathrm{ff}}$ at the center when the filamentary cloud fragments, pressure gradient force is important to calculate the further evolution of fragments. Further evolution of each fragment is shown in $\S 4$.

\subsubsection{Case with the external radiation}
\begin{figure}
\begin{center}
\begin{tabular}{cc}
\resizebox{80mm}{!}{\includegraphics{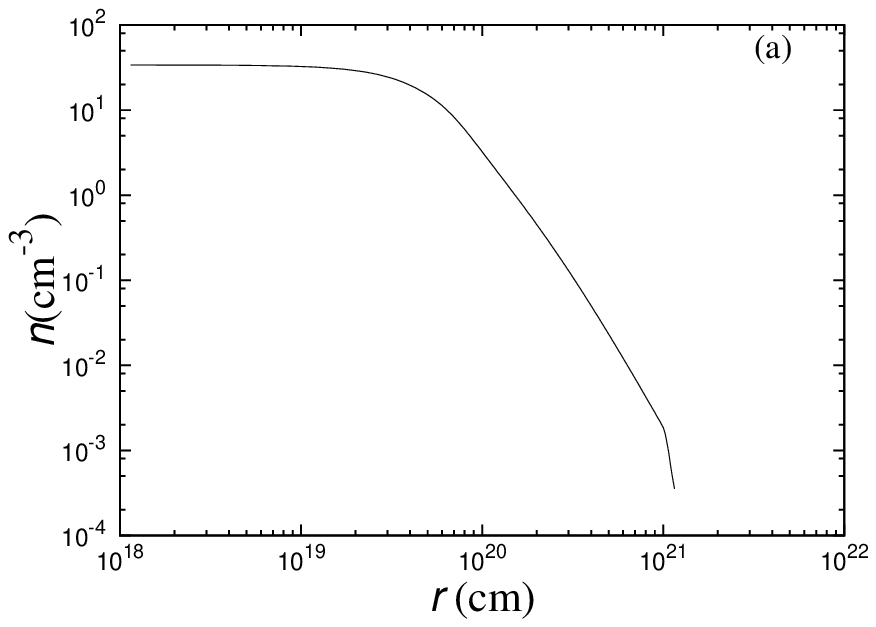}} &
\resizebox{80mm}{!}{\includegraphics{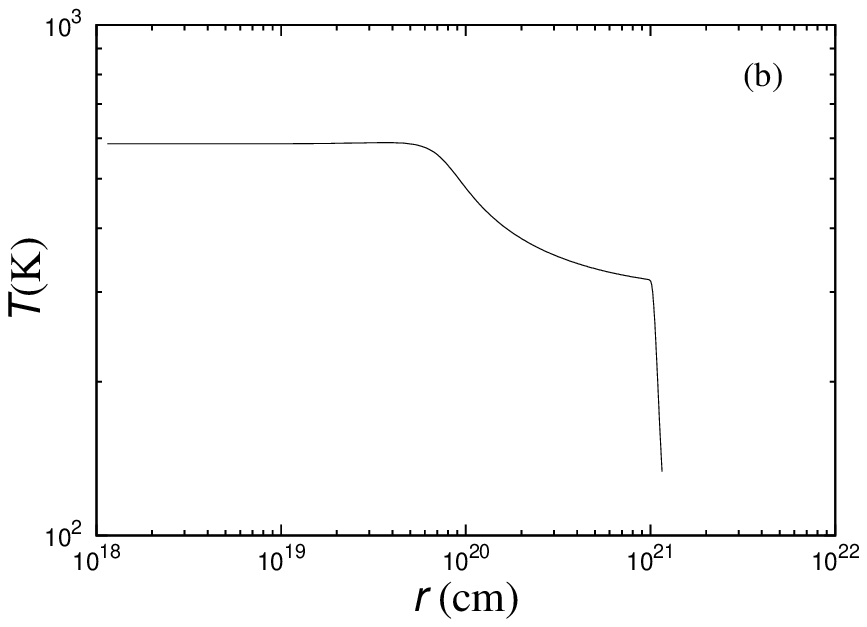}} \\
\resizebox{80mm}{!}{\includegraphics{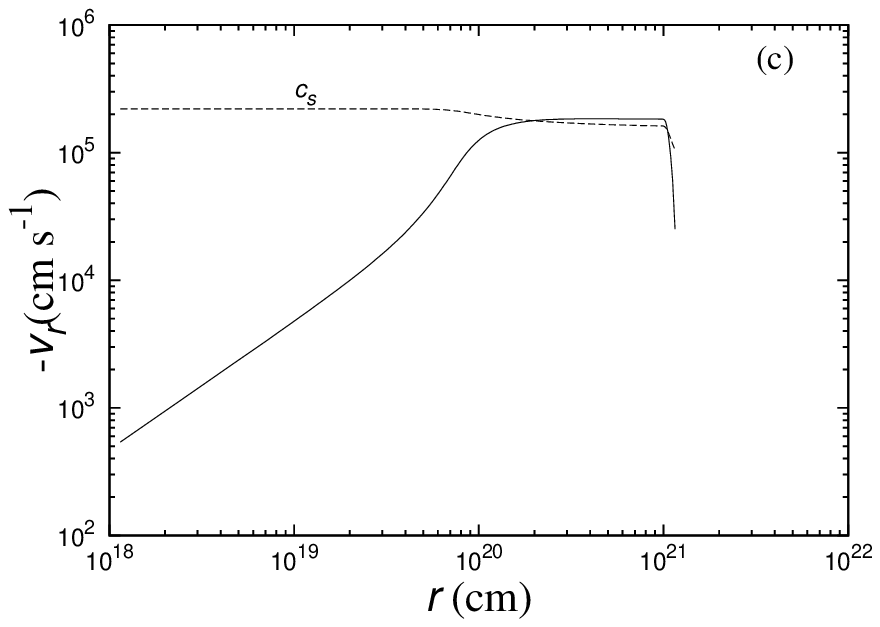}} &
\resizebox{80mm}{!}{\includegraphics{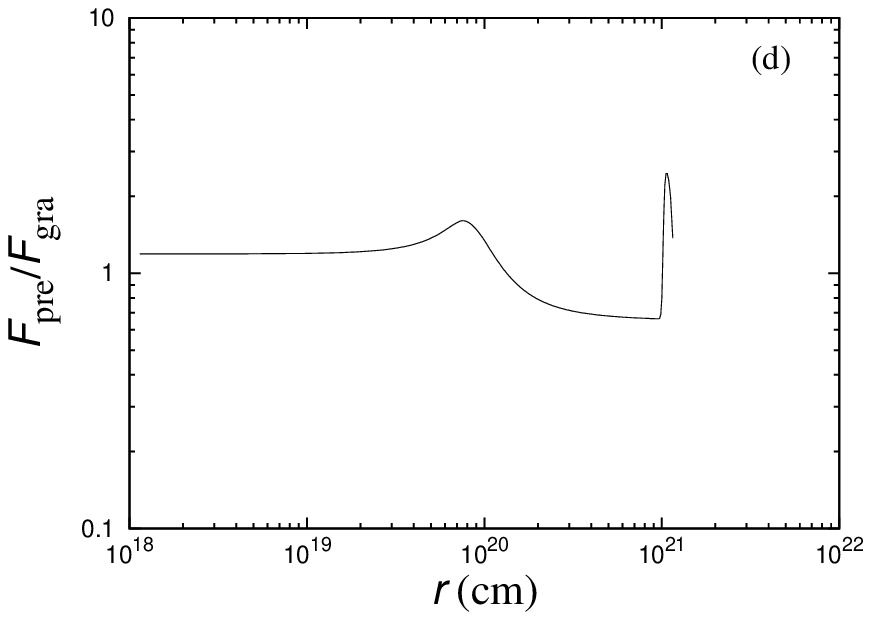}} \\
\end{tabular}
\caption{Same as figure 8, but $(f,n_0,J_{21})=(1.5,10\mathrm{cm^{-3}},10)$.}
\end{center}
\end{figure}

We show profiles of the same quantities as figure 8 for the case with $(f,n_0,J_{21})=(1.5,10\mathrm{cm^{-3}},10)$ in figure 9. In figure 9, it is seen that except for temperature, profiles of physical quantities are similar to figure 8. Most of $\mathrm{H_2}$ is photodissociated, and the filamentary cloud loses the ability to cool. Hence, temperature is higher in the central dense region than in the outer envelope. Ratio of pressure gradient to gravity is larger than $1$ ($\sim 1.2$). Fragment mass is $2.4 \times 10^5 M_\odot$ which is close to Jeans mass ($3.0 \times 10^5 M_\odot$) estimated with central density and temperature. In the case with the external radiation, since $t_{\mathrm{dyn}}$ is about 5 times of $t_{\mathrm{ff}}$ at the center when the filamentary cloud fragments, pressure gradient force is important to calculate the further evolution of fragments (see $\S 4$). In figure 10, the density profiles in figure 8 and 9 are simultaneously plotted, and each profile is found to be similar to each other. The profiles at $r/\lambda_J<0.5$ are similar to the profile of isothermal filamentary cloud in equilibrium state.

\begin{figure}
\begin{center}
\begin{tabular}{cc}
\resizebox{80mm}{!}{\includegraphics{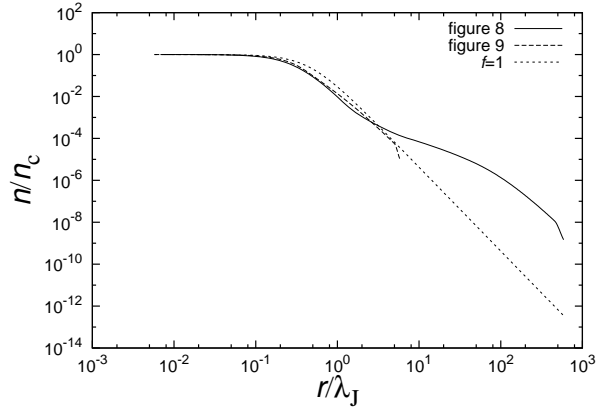}} \\
\end{tabular}
\caption{Normalized density profile for figure 8 (solid line) and figure 9 (dashed line). The initial density profile in the case with $f=1$ is also shown (dotted line). The symbol $n_c$ is central density and $\lambda_J$ is Jeans length at the center.}
\end{center}
\end{figure}

\subsection{Fragment mass}
\begin{figure}
\begin{center}
\begin{tabular}{cc}
\resizebox{80mm}{!}{\includegraphics{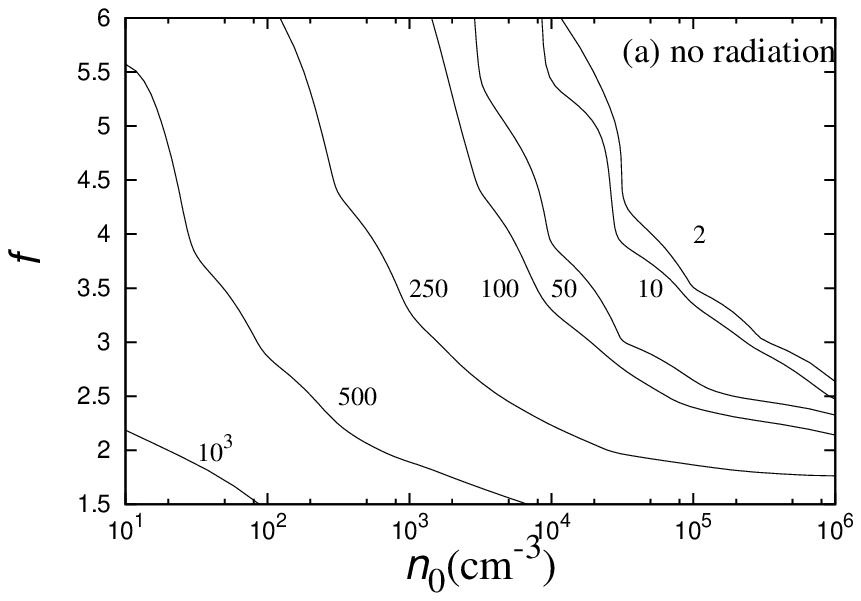}} &
\resizebox{80mm}{!}{\includegraphics{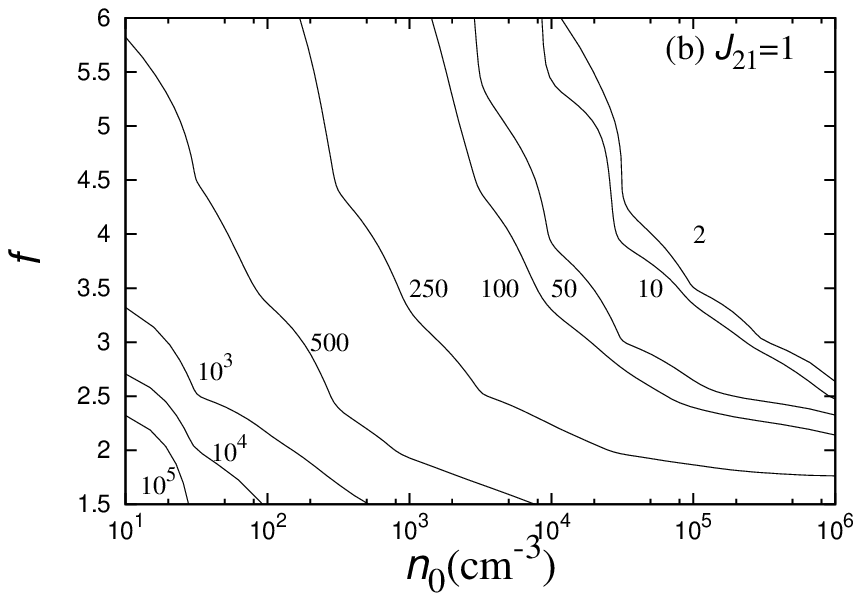}} \\
\resizebox{80mm}{!}{\includegraphics{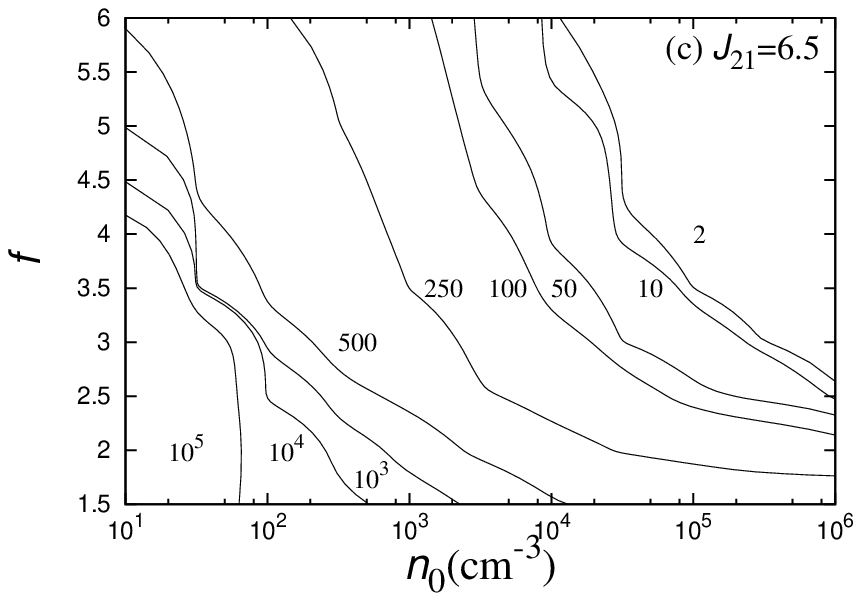}} &
\resizebox{80mm}{!}{\includegraphics{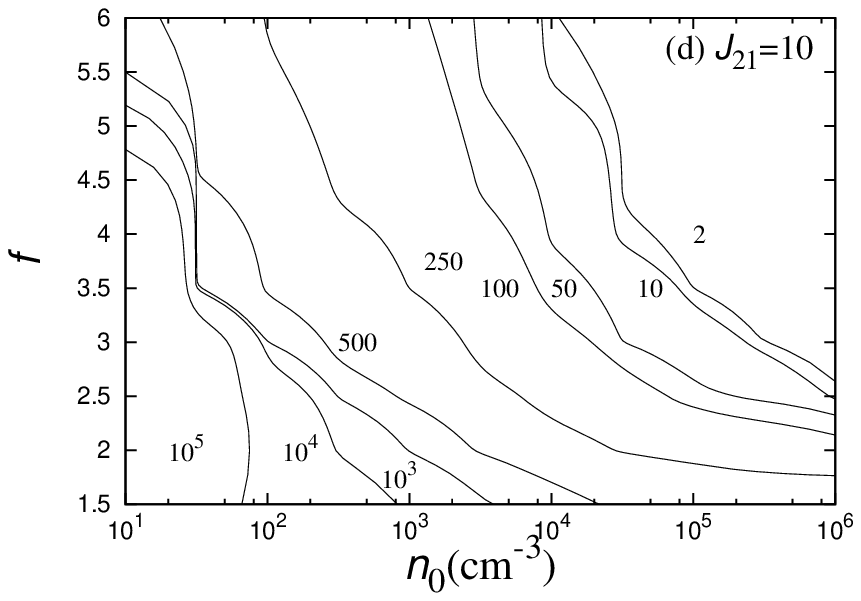}} \\
\end{tabular}
\caption{The contours map for the fragment mass for the case with (a) $J_{21}=0$, (b) $J_{21}=1$, (c) $J_{21}=6.5$, and (d) $J_{21}=10$. Solid lines in each diagrams represent constant fragment mass. The number near each solid line is mass of fragment in units of $M_\odot$.}
\end{center}
\end{figure}

We show how much the external dissociation radiation changes fragment mass. Figure 11 shows the fragment mass for all the parameters in $n_0 - f$ plane using contours. Results for the cases with $J_{21}=0$, $1$, $6.5$, and $10$ are presented in different diagrams. In the case with the external radiation, it is seen that the filamentary clouds fragment into very massive clouds ($> 10^5 M_\odot$) in the cases with low initial density ($n_0 \le 10^2 \mathrm{cm^{-3}}$). Since very massive fragments are not seen in the case without the external radiation, this can be regarded as a result of the effect of the dissociation photon. This feature is similar to the result of the one-zone models in Paper I. Thus, formation of very massive fragments under the external radiation with moderate intensity can be regarded as the robust result, provided that the external radiation turns on when the filamentary cloud forms.

The diagram (a) of figure 11 is similar to figure 6 of Nakamura $\&$ Umemura (2001). In the range of $2-100M_\odot$, the contours are dense. This is because the filamentary cloud becomes isothermal once $\mathrm{H_2}$ cooling becomes effective owing to three body reaction, and continues to collapse to high density ($\sim 10^{13} \mathrm{cm^{-3}}$). In such a case, fragment mass is small ($\sim 2-10 M_\odot$). This feature is seen in Nakamura $\&$ Umemura (2001).

Nakamura $\&$ Umemura (2002) concluded that there are some parameter sets where $\mathrm{HD}$ is main coolant. However, in our results, $\mathrm{HD}$ is found not to be important. Deuterated hydrogen molecules $\mathrm{HD}$ mainly forms from $\mathrm{H_2}$\footnote{Main chemical reaction of formation of $\mathrm{HD}$ is given by
\begin{eqnarray}
\mathrm{H_2} + \mathrm{D^+} \rightarrow \mathrm{HD} + \mathrm{H^+}.
\end{eqnarray}
}. Since initial $\mathrm{H_2}$ fraction in this paper is assumed to be small ($10^{-4}$), even in the case without the external radiation, sufficient amount of $\mathrm{HD}$ to cool does not form. This result is consistent with Nakamura $\&$ Umemura (2002). In the case with the external radiation, since $\mathrm{H_2}$ is photodissociated, $\mathrm{HD}$ is less important.

\begin{figure}
\begin{center}
\begin{tabular}{cc}
\resizebox{80mm}{!}{\includegraphics{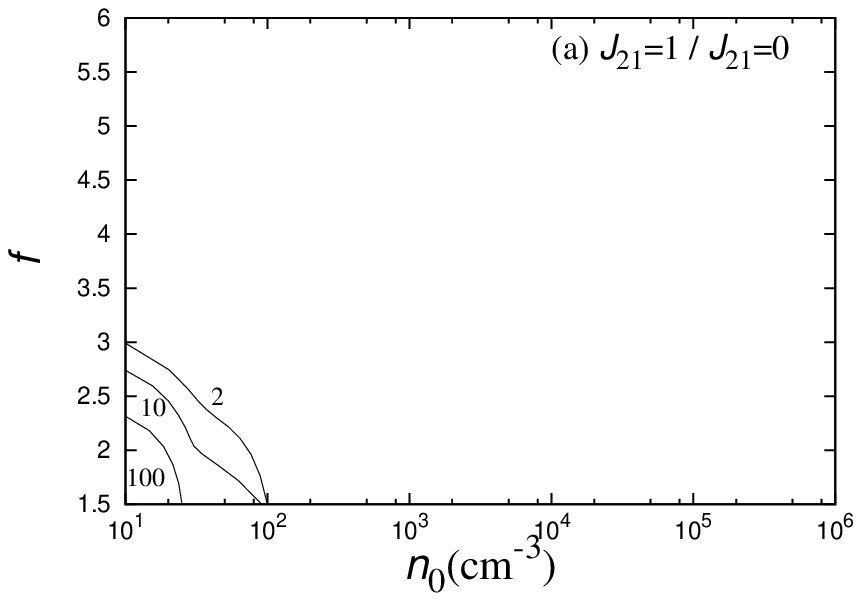}} &
\resizebox{80mm}{!}{\includegraphics{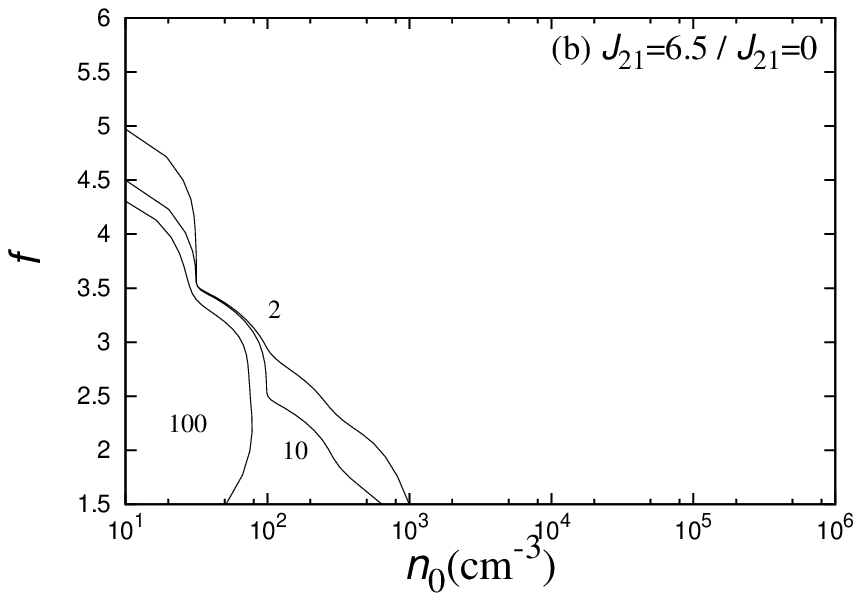}} \\
\resizebox{80mm}{!}{\includegraphics{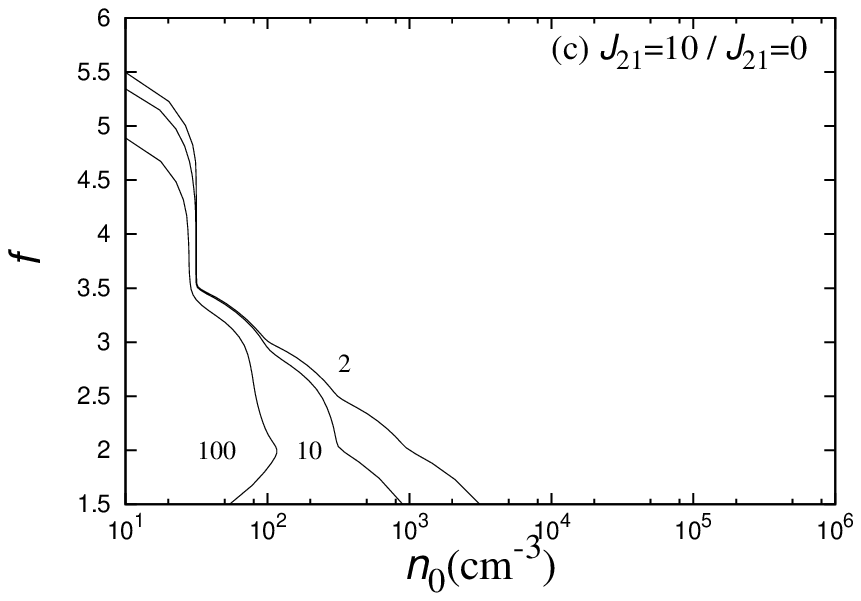}}
\end{tabular}
\caption{The ratio of the fragment mass with the external radiation ((a) $J_{21}=1$, (b) $J_{21}=6.5$, and (c) $J_{21}=10$) to that without the external radiation ($J_{21}=0$). The number near each solid line is the ratio of fragment mass.}
\end{center}
\end{figure}

We show the effect of the external radiation on fragment mass quantitatively. Figure 12 shows the similar contours but about the ratio of the fragment mass between the cases with and without the external radiation. In addition to figure 11, figure 12 clearly shows that the filamentary clouds with low initial density ($n_0 \le 10^2 \mathrm{cm^{-3}}$) and moderate $f$ ($<4.5$) fragment into more massive fragments than the case without the external radiation. This feature agrees with the results of the rarefied filament model in Paper I.

It is seen that fragment mass for the case with high initial density ($n_0 > 10^2 \mathrm{cm^{-3}}$) does not increase owing to the external dissociation radiation. This is because the initial density is high enough for the filamentary cloud to shield itself from the external dissociation radiation. In this case, the evolution is similar to the case without the external radiation.

\subsection{Self-shielding function of Wolcott-Green et al. (2011)}
Wolcott-Green et al. (2011) have recently suggested self-shielding function indicated by three-dimensional radiative transfer. In this subsection, we investigate dependence of fragment mass on the self-shielding function quantitatively. The new self-shielding function is given by,
\begin{eqnarray}
f_{\mathrm{sh,WG}}=\frac{0.965}{(1+x/b_5)^{1.1}}+\frac{0.035}{(1+x)^{0.5}} \exp [-8.5 \times 10^{-4} (1+x)^{0.5}],
\end{eqnarray}
where
\begin{eqnarray}
x \equiv \frac{N_{\mathrm{H_2}}}{5 \times 10^{14} \mathrm{cm^{-2}}},
\end{eqnarray}
and
\begin{eqnarray}
b_5 \equiv \frac{b}{10^5 \mathrm{cm/s}} = \frac{1}{10^5 \mathrm{cm/s}} \sqrt{\frac{2k_B T}{\mu m_{\mathrm{H}}}}.
\end{eqnarray}
Since $f_{\mathrm{sh,WG}}$ is $1-10$ times larger than original $f_{\mathrm{sh}}$ (equation 12). Hence, Wolcott-Green et al. (2011) suggested that the effect of photodissociation is actually stronger than the result with $f_{\mathrm{sh}}$. Thus, the results which we have ever considered are expected to be modified quantitatively.

\begin{figure}
\begin{center}
\begin{tabular}{cc}
\resizebox{80mm}{!}{\includegraphics{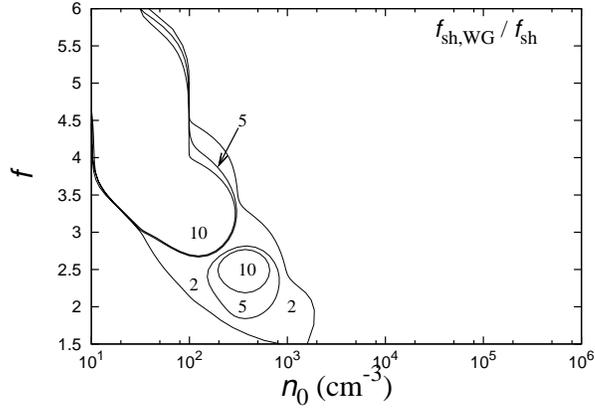}} \\
\end{tabular}
\caption{The ratio of the fragment mass with $f_{\mathrm{sh,GW}}$ to that with $f_{\mathrm{sh}}$ in the case with $J_{21}=10$. The number near each solid line is the ratio of fragment mass.}
\end{center}
\end{figure}

To see the difference between shielding function, we calculate fragment mass using $f_{\mathrm{sh,WG}}$ in the case with $J_{21}=10$. Figure 13 shows the ratio of fragment mass between the cases with $f_{\mathrm{sh,GW}}$ and with $f_{\mathrm{sh}}$. When we use $f_{\mathrm{sh,GW}}$, fragment mass increases comparing with the case with $f_{\mathrm{sh}}$ especially for the cases with $n_0=10-10^3 \mathrm{cm^{-3}}$ and $f \ge 2$. This feature is consistent with the relation, $f_{\mathrm{sh,GW}} \ge f_{\mathrm{sh}}$.

\section{Criterion for increase of fragment mass}
Omukai $\&$ Yoshii (2003) calculated the evolution of the filamentary cloud under the external dissociation radiation assuming free-fall. The authors assumed that fragmentation occurs at density 100 times higher than the loitering point and concluded that the effect of the external dissociation radiation decreases fragment mass. This conclusion apparently disagrees with our results in $\S 3$. In this section, we consider whether or not the external radiation increases fragment mass when the filamentary cloud reaches the loitering point. Initial density of the filamentary cloud is assumed to be very low ($n_0=0.1 \mathrm{cm^{-3}}$). As the timing when the external radiation turns on, we consider various cases. The investigation in this section provides systematic study which includes the situation of Omukai $\&$ Yoshii (2003).

\subsection{Whether the filament reaches the loitering point}
Suppose a filamentary cloud with $n_0=0.1 \mathrm{cm^{-3}}$ and the external radiation turns on when density reaches $n_{\mathrm{UV}}$. We consider the cases with $n_{\mathrm{UV}}=0.1\mathrm{cm^{-3}}$, $1\mathrm{cm^{-3}}$, and $10\mathrm{cm^{-3}}$. We also assume $T_0=300\mathrm{K}$ and $f_{\mathrm{H_2}}=0$ at $n_0$. As for the evolution of the filamentary cloud, we solve one-dimensional hydrodynamics as in $\S 3$.

We calculate the evolution of the filamentary cloud with various values of $f$. In order for the filamentary cloud not to fragment during adiabatic phase in the case with $J_{21}=10$, it is found that $f$ is required to be larger than $30$, $25$, $10$, and $5$ for various values of $n_{\mathrm{UV}}$ ; $n_{\mathrm{UV}}=0.1\mathrm{cm^{-3}}$, $1 \mathrm{cm^{-3}}$, $10\mathrm{cm^{-3}}$, and $\infty$, respectively. Hence, we investigate how massive fragments are in the case with $f=30$, $25$, $10$, and $5$. Does fragment mass increase when the filamentary cloud reaches the loitering point?

\begin{figure}
\begin{center}
\begin{tabular}{cc}
\resizebox{80mm}{!}{\includegraphics{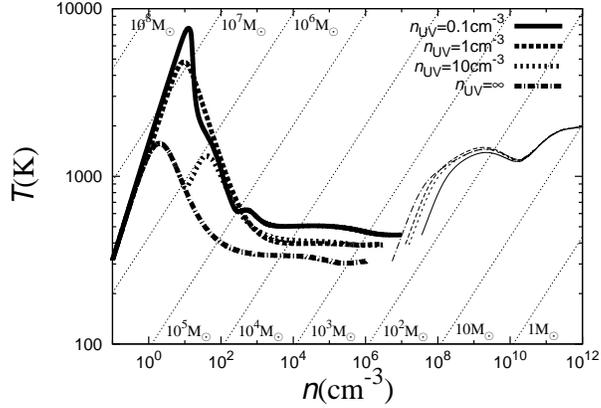}} \\
\end{tabular}
\caption{Thermal evolution of the filamentary cloud with $f=30$, $n_0=0.1\mathrm{cm^{-3}}$, $J_{21}=10$, and various $n_{\mathrm{UV}}$ where the external radiation turns on ; $n_{\mathrm{UV}}=0.1\mathrm{cm^{-3}}$ (thick solid line) $n_{\mathrm{UV}}=1\mathrm{cm^{-3}}$ (thick long-dashed line), $n_{\mathrm{UV}}=10\mathrm{cm^{-3}}$ (thick short-dashed line), $n_{\mathrm{UV}}=\infty$ (thick dash-dotted line). Thin lines shows the thermal evolution of fragments. Dotted lines indicate the constant Jeans masses.}
\end{center}
\end{figure}

First, we show the case with $f=30$ where the filamentary cloud reaches the loitering point for any $n_{\mathrm{UV}}$. Figure 14 shows the thermal evolution of the filamentary cloud with $f=30$, $n_0=0.1\mathrm{cm^{-3}}$, $J_{21}=10$, and various $n_{\mathrm{UV}}$. Thin lines indicate thermal evolution of each fragment (see $\S 4.2$). Fragment mass is largest ($550M_\odot$) in the case with $n_{\mathrm{UV}}=\infty$. Fragment mass is smaller in the case with lower $n_{\mathrm{UV}}$. As a result, the external radiation decreases fragment mass. All filamentary clouds fragments at the density which is roughly $1000$ times higher than loitering point, which is qualitatively consistent with Omukai $\&$ Yoshii (2003).

\begin{figure}
\begin{center}
\begin{tabular}{cc}
\resizebox{80mm}{!}{\includegraphics{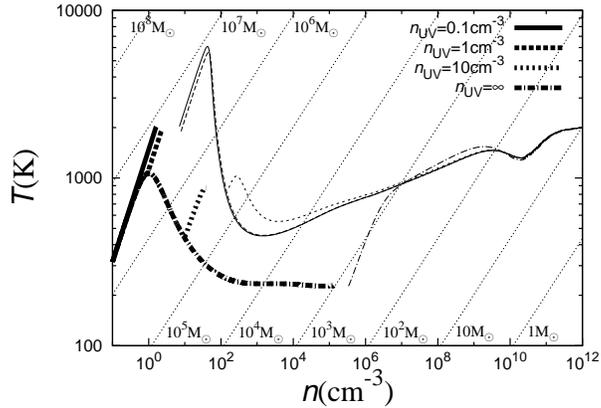}} \\
\end{tabular}
\caption{Same as figure 14, but $f=5$.}
\end{center}
\end{figure}

Next, we investigate whether or not the external radiation decreases fragment mass when the filamentary cloud does not reach the loitering point. Figure 15 shows the same figure as figure 14 except for the value of $f$ : $f=5$. It is found that the filamentary cloud which does not reach the loitering point fragments into more massive clouds than the case with $n_{\mathrm{UV}}=\infty$. This feature agrees with our result of $\S 3$. Also in the cases with $f=25$ and $10$, the filamentary cloud fragments into more massive clouds than the case without the external radiation ($n_{\mathrm{UV}}=\infty$) when it does not reaches loitering point ($n_{\mathrm{UV}} \le 0.1 \mathrm{cm^{-3}}$ for $f=25$ and $n_{\mathrm{UV}} \le 1 \mathrm{cm^{-3}}$ for $f=10$). Whether or not the external radiation increases fragment mass is determined by whether or not line mass is small enough for the filamentary cloud not to reach the loitering point.

It is explained as follows that fragment mass decreases under the external radiation : after the external radiation turns on, the filamentary cloud suffers photodissociation. The outer region loses more amount of $\mathrm{H_2}$ than the dense central region. Since temperature of the outer region becomes higher than the dense central region, pressure gradient becomes sallower at the region between the dense central region and the outer region. Then, such a region sticks to the dense central region owing to gravity, and line mass of the central dense region increases. As a result, the filamentary cloud collapses up to higher density ($n \sim 10^{6-7} \mathrm{cm^{-3}}$) owing to larger line mass and fragments into less massive clouds than the case without the external radiation.

\subsection{Sub-fragmentation}
It is found that the external radiation increases fragment mass when the filamentary cloud fragments before the loitering point. However, each fragment may occur sub-fragmentation and mass of final outcome may be as small as fragment mass in the case without the external radiation. To clarify mass of final outcome, we investigate the further evolution of each fragment.

We use one-zone model for each fragment. We assume that each fragment is spherical and has $M_{\mathrm{frag}}$ (see equation 14), chemical composition same as at the center of the filamentary cloud, radius $r_{0.1}$ where $n (r_{0.1})=0.1 n_c$. Infall velocity at fragmentation is proportional to radius, and we use infall velocity at $r_{0.1}$ as infall velocity of each fragment. The density of fragment is estimated from the mass and radius as
\begin{eqnarray}
\rho_{\mathrm{frag}}=\frac{M_{\mathrm{frag}}}{\frac{4\pi}{3} r_{0.1}^3}.
\end{eqnarray}
This density, $\rho_{\mathrm{frag}}$, is different from the density of the filamentary cloud before fragmentation and corresponds to the one at the end of fragmentation. Temperature and fraction of each fragment is approximated by the value when the filamentary cloud fragments. According to the result in $\S 3.3$, in the view point of dynamical evolution, we should consider the effect of pressure gradient force. Hence, we treat pressure effect explicitly using the virial equation for uniform sphere. The virial equation is given by
\begin{eqnarray}
\frac{dv}{dt}=\frac{10}{3}\frac{k_B T}{\mu m_{\mathrm{H}}} \frac{1}{R} -\frac{GM}{R^2},
\end{eqnarray}
where $R$ is radius and $M$ is mass of the cloud (see Appendix 1). Radiative transfer is treated by the same method as in Omukai (2001) except for we use $f_{\mathrm{sh,GW}}$ instead of $f_{\mathrm{sh}}$. The same routine as $\S 3$ is used for chemical reactions.

In figure 15, it is found that if each fragment occurs sub-fragmentation at the loitering point ($n \sim 10^3 \mathrm{cm^{-3}}$), mass of final outcome is $\sim 10^4 M_\odot$. This is larger than fragment mass in the case without the external radiation. Hence, even if sub-fragmentation occurs, mass of final outcome increases owing to the external radiation when the filamentary cloud fragments before the loitering point. This tendency is found in the other cases with $f=25$ and $10$.

When the filamentary cloud reaches the loitering point, temperature of each fragment increases adiabatically after fragmentation ($n \sim 10^{6-7} \mathrm{cm^{-3}}$). This is because gravity dominates pressure gradient after fragmentation and dynamical time becomes shorter than cooling time.

\section{Conclusions and discussions}
In this paper, collapse and fragmentation of primordial filamentary cloud was investigated using one-dimensional hydrodynamical calculations with the effect of the external dissociation radiation. Especially, the effect of run-away collapse to fragment mass is considered by comparing with previous results with one-zone models. Results are summarized as follows :
\begin{itemize}
\item Comparing with the uniform model in Paper I, one-dimensional filament model predicts lower fragmentation density and larger fragment mass. This is because fragmentation occurs only in the central region with low virial temperature in one-dimensional model.
\item Comparing with the rarefied filament model in Paper I, one-dimensional filament model predicts similar fragment mass. This explains that the discrepancy between the uniform filament model and one-dimensional filament model mainly comes from the run-away collapse which is partly induced by the pressure effect.
\item As long as the external radiation is assumed to turn on when the filamentary clouds form, low initial density ($n_0 \le 10^2 \mathrm{cm^{-3}}$) filamentary clouds with moderate line mass are expected to fragment into very massive clumps ($\sim 10^5 M_\odot$) as a result of photodissociation of molecular hydrogen. This result which is originally indicated in Paper I is confirmed in this paper using one-dimensional hydrodynamical calculations.
\item The external dissociation radiation increases fragment mass when the filamentary cloud fragments during adiabatic phase after the external radiation turns on. On the other hand, when the filamentary cloud with sufficient line mass reaches the loitering point, the dissociation radiation decreases fragment mass, which is consistent with Omukai $\&$ Yoshii (2003).
\end{itemize}

As seen in figure 15, the thermal evolution of a filamentary cloud and a spherical cloud is different after each cloud reaches loitering point. The thermal evolution of the filamentary cloud is isothermal, and temperature of spherical cloud increases. This is explained as follows : since the central region of the filamentary cloud is approximately dynamical equilibrium ($\S 3.3$), we have
\begin{eqnarray}
\frac{1}{\rho} \frac{P}{r} &\sim& \frac{Gl}{r} \nonumber \\
T &\sim& \mu m_{\mathrm{H}} Gl \propto const.
\end{eqnarray}
As for spherical cloud, the central region is not dynamical equilibrium. Hence, we investigate relation between $T$ and $n$ from balance between adiabatic heating and $\mathrm{H_2}$ cooling. Adiabatic heating rate is
\begin{eqnarray}
-P \frac{d\ }{dt} \frac{1}{\rho} \sim P \frac{1}{t_{\mathrm{ff}}\rho} \propto T n^{1/2}
\end{eqnarray}
where we assume that spherical cloud collapses in free-fall timescale. When density is larger than the critical density of $\mathrm{H_2}$, $\Lambda_{\mathrm{H_2}} \propto n T^\alpha$ ($\alpha \sim 3.8$ at $T \sim 300 \mathrm{K}$ and $\alpha \sim 4.8$ at $T \sim 1000 \mathrm{K}$), and we have
\begin{eqnarray}
\frac{\Lambda_{\mathrm{H_2}}}{\rho} &\propto& \frac{n T^\alpha}{n} \propto T^\alpha.
\end{eqnarray}
From equations (29) and (30), we have $T \propto \rho^{1/2(\alpha-1)}$. Temperature depends on density as,
\begin{eqnarray}
T \propto \left\{ \begin{array}{ll}
n^{1/2(\alpha-1)} & \mathrm{sphere} \\
const. & \mathrm{filament}. \\
\end{array} \right.
\end{eqnarray}

In this paper, we consider the filamentary cloud with variety of line mass. Hence, we estimate $f$ for the filamentary clouds in cosmological simulation. As an example, we refer with figure 2 of Greif et al. (2008). In this figure, the filamentary cloud with density $n \sim 10^{-2} \mathrm{cm^{-3}}$ and radius $\sim 7 \mathrm{kpc}$ is shown. Line mass of this filamentary cloud is $\sim 7.8 \times 10^{18} \mathrm{g/cm}$. When temperature is $300\mathrm{K}$, the critical line mass is $l_{\mathrm{crit}} \sim 3.5 \times 10^{17} \mathrm{g/cm}$. Hence, $f \sim 22$ and according to our results in $\S 4$, fragment mass may increase if the external radiation turns on at $n \le 1 \mathrm{cm^{-3}}$.

For simplicity, the model and numerical calculations in this paper are one-dimensional for the filamentary cloud and one-zone for each fragment. In order to discuss fragmentation, we assume the condition for fragmentation ($\S 2.3$) and assume that each fragment is spherical. During further collapse, spherical clouds are possible to be the filamentary cloud again or may become disk-like if it rotates. Although we discussed a possible sub-fragmentation at the loitering point for each clumps in $\S 4.2$, the final fate of the cloud is still open question. Furthermore, if the filamentary cloud with large line mass is not axisymmetric, it may become sheet-like cloud. Such a cloud may collapse and fragment into many filamentary clouds. In this paper, the external radiation is assumed to be uniform, and the intensity does not depend on time. These problems require three-dimensional calculations. As for three-dimensional simulation with the external radiation, Susa (2007) investigated collapse of spherical cloud under the single light source. However, further investigations with three-dimensional simulation which statistically investigate fragment mass of the filamentary cloud under the external radiation will be desirable. Despite simplicity, one-dimensional hydrodynamical calculations in this paper are useful in the view point of extracting physical processes which are important in formation of the astronomical objects. These one-dimensional calculations and the realistic three-dimensional calculations may be complementary.

\bigskip

We thank Fumio Takahara for fruitful discussion and continuous encouragement. We also acknowledge the referee for improving the manuscript.

\appendix
\section{Virial equation for uniform sphere}
We multiply $4 \pi r^3$ by both hands of equation of motion,
\begin{eqnarray}
\rho \frac{Dv}{Dt}=-\frac{dP}{dr}-\frac{GM}{r^2}\rho,
\end{eqnarray}
and integrate in respect with $r$. Then, the left side hand of equation (A1) is
\begin{eqnarray}
\int 4\pi r^3 \rho \frac{Dv}{Dt} dr &=& \int r \frac{Dv}{Dt} dM \nonumber \\
&=& \int \biggl( \frac{1}{2} \frac{D^2\ }{Dt^2} r^2 - v^2 \biggr) dM \nonumber \\
&=& \frac{1}{2}\frac{D^2\ }{Dt^2} \int r^2 dM - \int v^2 dM.
\end{eqnarray}
About the first term of right side hand of equation (A2),
\begin{eqnarray}
\int r^2 dM &=& \int 4 \pi r^4 \rho dr \nonumber \\
&=& \frac{4\pi \rho}{5} R^5 \nonumber \\
&=& \frac{3}{5}MR^2,
\end{eqnarray}
and 
\begin{eqnarray}
\frac{1}{2}\frac{D^2\ }{Dt^2} \int r^2 dM &=& \frac{1}{2} \frac{D^2\ }{Dt^2} \biggl( \frac{3}{5} MR^2 \biggr) \nonumber \\
&=& \frac{3}{5}M \biggl(\frac{DR}{Dt} \biggr)^2 + \frac{3}{5}MR \frac{D^2R}{Dt^2}.
\end{eqnarray}
About the second term,
\begin{eqnarray}
\int v^2 dM &=& \int 4\pi r^2 \rho v^2 dr \nonumber \\
&=& \int 4 \pi \rho r^2 \biggl(\frac{DR}{Dt} \biggr)^2 \frac{r^2}{R^2} dr \nonumber \\
&=& \frac{3}{5} M \biggl(\frac{DR}{Dt} \biggr)^2,
\end{eqnarray}
where we use the following relation,
\begin{eqnarray}
v=\biggl(\frac{DR}{Dt}\biggr)\frac{r}{R},
\end{eqnarray}
since velocity is in proportion to $r$ because of uniform density. Hence, the left side hand of equation (A2) is
\begin{eqnarray}
\int 4\pi r^3 \rho \frac{Dv}{Dt} dr = \frac{3}{5} MR \frac{D^2\ }{Dt^2} R.
\end{eqnarray}
On the other hand, about the right side hand of equation of motion, term of pressure gradient is
\begin{eqnarray}
- \int 4 \pi r^3 \frac{dP}{dr} dr = 3(\gamma_{\mathrm{adi}}-1) \frac{k_B T}{\mu m_{\mathrm{H}}} M,
\end{eqnarray}
where $\gamma_{\mathrm{adi}}$ is adiabatic index. The second term is
\begin{eqnarray}
- \int 4\pi r^3 \frac{GM}{r^2} \rho dr &=& \int (4 \pi \rho)^2 \frac{G}{3}r^4 dr \nonumber \\
&=& -\frac{3}{5}\frac{GM^2}{R}.
\end{eqnarray}
Finally, we have virial equation for uniform sphere,
\begin{eqnarray}
\frac{3}{5}MR\frac{Dv}{Dt}&=&3(\gamma_{\mathrm{adi}}-1)\frac{k_B T}{\mu m_{\mathrm{H}}} \frac{1}{R} - \frac{3}{5}\frac{GM}{R} \\
\frac{Dv}{Dt}&=&\frac{10}{3}\frac{k_B T}{\mu m_{\mathrm{H}}} \frac{1}{R} -\frac{GM}{R^2},
\end{eqnarray}
where we use $\gamma_{\mathrm{adi}}=5/3$.


\end{document}